# Materials under XUV irradiation: effects of structure, size, and temperature

Nikita Medvedev*[a,b]

[a] Institute of Physics, Czech Academy of Sciences, Na Slovance 1999/2, 18221 Prague 8, Czechia;
[b] Institute of Plasma Physics, Czech Academy of Sciences, Za Slovankou 3, 18200 Prague 8, Czechia;

## ABSTRACT

This proceeding discusses the impact of XUV/X-ray irradiation on materials, and how their response is affected by temperature, size, and structure. When materials are exposed to intense XUV/X-ray irradiation, they undergo a series of processes ultimately leading to observable structure modification and damage. These effects were studied with a hybrid simulation tool XTANT-3. The code combines several methods in one interconnected model: the photon absorption and electron cascades are simulated with transport Monte Carlo; nonequilibrium kinetics of slow electrons (in the valence and the bottom of the conduction band) is traced with the Boltzmann equation; modeling evolution of the electronic structure and interatomic potential is done with the transferable tight binding method; and the response of the atomic system is simulated with the molecular dynamics. Combining these methods enabled the tracing of the essential effects of irradiation. This brief review summarizes the recent results obtained with this simulation tool.

## 1. INTRODUCTION

Since the development of free-electron lasers, the intense ultrashort femtosecond XUV and X-ray pulses became available for studies of materials response to such irradiation in a controlled way [1–3]. It opened up possibilities for studies of transient nonequilibrium states of matter produced by irradiation at their natural timescales. Irradiation of matter with ultrashort laser pulses plays a key role in basic and applied sciences [4–8]. Laser irradiation experiments are used for material processing, nano-, micro-technology, medical applications such as laser surgery, etc.[7–10]. Femtosecond X-ray irradiation allows for subatomic precision measurements of atomic and electronic dynamics [1–3].

X-ray pulse impinging on a material surface triggers a sequence of processes. It starts with photoabsorption by core atomic shells, exciting electrons to high-energy states [11]. Such electrons then form an electron cascade, exciting secondary electrons *via* impact ionization of core shells and valence/conduction band of the material, and transferring energy to the target atoms *via* elastic channel (electron-ion or electron-phonon scattering) [12,13]. Created core-shell holes decay primarily *via* the Auger channel (for the photon energies relevant for currently existing X-ray free-electron lasers), thereby exciting secondary electrons [14].

Electrons, relaxed to the bottom of the conduction and the valence band, influence the interatomic potential. The interatomic potential changes its shape due to the excitation of electrons from their ground state, which may destabilize the atomic lattice and induce a phase transition even without heating the lattice to the melting point. As soon as electrons are excited, atoms start to react to the modification of the interatomic potential, which in the case of femtosecond lasers may lead to the ultrafast phase transition. The effect is known as the nonthermal melting (or, more generally, nonthermal phase transition) [15,16].

At longer times, electron-ion (electron-phonon) coupling heats the atomic system, which may induce thermal phase transition – such as melting or solid-solid phase transition. This transition takes place when heated atoms overcome their potential energy barrier and end up in a new phase. The barriers, being defined by the interatomic potential energy surface, are also affected by electronic excitation. Thus, a nontrivial interplay between thermal and nonthermal effects may take place in irradiated matter [17–19].

* Corresponding author: ORCID: 0000-0003-0491-1090, email: nikita.medvedev@fzu.cz

All the above-mentioned processes are sensitive to some degree to the particular state of the material before irradiation. The material structure, the shape of the sample, and the irradiation temperature (in situ temperature before the arrival of the laser pulse) – all affect the material response. Here, we summarize our recent theoretical research on the influence of these factors on the material response to irradiation.

## 2. OUTLINE OF THE MODEL: XTANT-3

To study the material irradiation with ultrashort X-ray pulses, we applied the hybrid code XTANT-3 [17,20]. The code combines on-the-fly the following modules, describing the effects and processes taking place in the irradiated sample, see Figure 1.

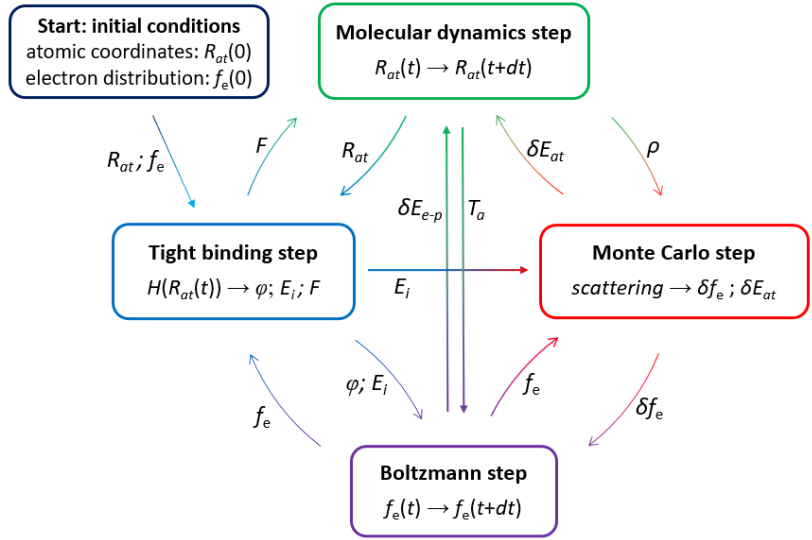

Figure 1. Schematics of the XTANT-3 algorithms, interconnecting the main modules of the program. The initial conditions specify the coordinates of all the atoms in the simulation box $R_{at}(t=0)$ and electronic distribution function $f_e$. The molecular dynamics (MD) module propagates the atomic coordinates, using the forces, $F$, calculated with the help of the tight binding (TB) module. It also uses the electron transfer from the electronic system, calculated in two steps: elastic scattering of a high-energy fraction of electrons within the Monte Carlo (MC) step, $\delta E_{at}$, and electron-ion (electron-phonon) coupling of low-energy electrons calculated with the Boltzmann equation (BE), $\delta E_{el-phon}$. TB also provides the BE module with the transient electronic energy levels (eigenstates of the Hamiltonian), $E_i$, and wave functions (eigenfunctions) $\varphi$. $E_i$ are also used in MC. MC calculates the source terms for BE associated with the photoabsorption, Auger decays, and electronic scattering: $\delta f_e$. MD step supplies the BE with the atomic temperature $T_a$, and the MC with the transient material density $\rho$ (if the supercell size is allowed to change in the simulation).

(i) The photoabsorption, secondary electronic cascades, and Auger decays are simulated with the event-by-event Monte Carlo (MC) scheme [17]. Electrons are traced until their energy falls below the threshold of ~10 eV, counted from the bottom of the conduction band (or the Fermi level for metals). When an electron loses its energy below the threshold, it joins the low-energy fraction of electrons, described below.

(ii) The kinetics of low-energy electrons, populating the valence and the conduction band, is described with the Boltzmann equation (BE) [21]. It includes the electron-electron scattering in the framework of the relaxation-time approximation [21], the electron-phonon (electron-ion) coupling [22] via so-called dynamical coupling (based on Tully's nonadiabatic coupling matrix elements [23]), and the source terms calculated from MC module above.

(iii) The transient electronic structure is calculated with help of the transferrable tight binding (TB) model. Depending on the materials studied, XTANT-3 may apply widely used parameterizations such as NRL [24] or DFTB [25].

The electronic structure, or molecular orbitals, defined as the eigenstates of the transient Hamiltonian, depend on the mutual positions of all the atoms in the simulation box and thus evolve in time with the atomic motion. It also allows us to calculate the potential energy surface for atoms.

(iv) Atomic trajectories are traced with help of the molecular dynamics (MD) simulation. The interatomic potential or forces acting on atoms are calculated with help of the TB method above. Thus, the transient electronic populations, defined with BE, directly affect the interatomic potential. Additionally, the energy transfer from electrons (elastic scattering of high-energy electrons in the MC module, and electron-ion coupling in BE module) is fed to atoms at each timestep via velocity scaling [17].

Combining the nonequilibrium electronic description, nonadiabatic coupling, and nonthermal evolution of the interatomic potential, allows us to trace the synergistic response of the irradiated matter.

## 3. OVERVIEW OF RESULTS

Most of XTANT-3 simulations to date were performed assuming infinitely fast electron thermalization: $\tau_{e-e} \to 0$, with $\tau_{e-e}$ being the electron-electron characteristic scattering time. In this approximation, the electronic distribution function adheres to the Fermi-Dirac distribution at all time, under the condition of the conserved energy and number of particles. The only exception was the most recent work, Ref. [21], in which effects of the nonequilibrium electronic distribution were studied in various classes of materials: metals, semiconductors and dielectrics. The electronic nonequilibrium seems to play the strongest role in dielectrics, since in metals and to some extent semiconductors electron-electron thermalization is typically very fast, on the timescale of ~10 fs for the excitation levels around the material damage threshold[4]. All the results discussed below assume instantaneous electronic thermalization.

A series of works using XTANT-3 to study various conditions in various materials under irradiation have recently been published: the dependence of the electron-phonon coupling parameters on the electronic temperature for metals was reported in Ref. [22]; its dependence on the atomic temperature was discussed in more detail in Ref. [26]; the difference between bulk and nano-sized metals resulting in the formation of the nonthermal melting in the nano-latter was presented in Ref. [27]; the effect of the shape and dimensionality of a nano sample on the electron-phonon coupling and nonthermal ablation was reported in Ref. [28]; the effect of *in situ* irradiation temperature of material damage threshold was shown in Ref. [29]. Here, we will briefly summarize the findings of these works, and draw some general conclusions.

Let us start with the discussion on temperature dependence. Material preheating, before irradiation, may influence further behavior after the arrival of the laser pulse. For example, the electron-phonon coupling seems to be dependent on the atomic temperature nearly linearly [22,26]. Thus, irradiation of a hot material will affect the kinetics of the electron-ion energy exchange, and the characteristic timescales of the ensuing processes.

Even for the same amount of energy deposited in the sample (by means of preheating and incoming laser pulse), there is a difference if the energy was deposited during the pulse, or at the macroscopic timescales before it. In the preheating scenario, the material has sufficient time to relax, including macroscopic effects such as hydrodynamic expansion, whereas during the ultrafast pulse the energy is deposited in the material that did not have time to relax and adjust to the new conditions [29]. It results in different behavior of the sample, which may affect the damage threshold. For instance, the threshold of nonthermal graphitization of diamond seems to be reduced with an increase in the irradiation temperature, whereas the ablation threshold in tungsten increases due to pressure dissipation in the preheated metal [29].

The atomic structure or the phase of the material seems to be influencing the electron-phonon coupling more at the low electronic temperatures, but not as strongly at high ones [22,28]. The effect is different from the pure effect of the density – the electron-phonon coupling is inversely proportional to the material density (as long as the lattice structure is preserved), but the dependence qualitatively changes once the atomic structure changes (e.g., due to destabilization under expansion) [22].

The expansion of nano-sized metals induced by the increase of the electronic pressure due to ultrafast irradiation may by itself lead to atomic destabilization [27]. In such a process, all the material parameters also change, e.g. phonon spectra and thusly electron-phonon coupling [27].

The dimensionality of the sample also changes its response to irradiation. In the simulation comparing a nano-sphere (effectively 0d-sample), nano-rod (1d), nano-layer (2d), and bulk (3d), the electronic heat capacity and the electron-phonon coupling parameter were mostly affected at low electronic temperatures but were very close to each other at high ones (above some 5000-7000 K) [28]. However, the difference between the samples was more prominent in nonthermal behavior: smaller samples reacted to electronic pressure increase easier, expanding and destabilizing the atomic lattice [28].

The lattice destabilization also depended on the particular shape of the nanoparticle. An octahedral gold nanoparticle is damaged faster than a cubic one [28]. This effect is expected to play a stronger role for smaller particles – as the ratio of surface atoms to the interior ones increases.

There is also an effect of the electronic nonequilibrium: the transient state of the electronic distribution forming during and after irradiation may be different from the equilibrium Fermi-Dirac distribution. The nonequilibrium electronic ensemble alters all the other characteristics: electron heat conductivity, electron-phonon coupling, and nonthermal changes in the interatomic potential [21]. The coupling parameter seems to be smaller in the nonequilibrium case in comparison to the equilibrium distribution. The nonthermal damage threshold may be altered in either way – an increase or decrease – depending on the particular shape of the distribution function [21]. This effect may be important for a multi-pulse irradiation scenario, in which a prior pulse may alter the electronic distribution before the arrival of the next one.

The fact that the kinetic of damage and the threshold are sensitive to particular conditions of the sample and the pulse complicate the analysis of experimental data. For example, the measured electron-phonon coupling parameter in a metal may differ noticeably in different experiments, see discussion and examples in Ref. [22]. Thus, when extracting the dependence on a particular parameter – such as coupling parameter dependence on the electronic temperature – it is very important to control for all other parameters: the atomic lattice temperature, material density, size, and shape, as well as the laser pulse parameters (photon energy, pulse duration, shape etc.).

## 4. CONCLUSIONS

The material properties affect its response to ultrafast irradiation. The *in-situ* temperature may change the damage threshold in a nontrivial way: the material preheating and the energy deposited by radiation do not always add up to the same damage threshold. Heating material up may lower or increase the damage threshold, depending on particular mechanisms of damage (thermal, nonthermal, defect formation, ablation). Material size may change its properties and the damage mechanisms: e.g., bulk metals exhibit phonon hardening under irradiation whereas nano-sized metallic particles show nonthermal expansion and destabilization of the lattice. The particular shape of the nanoparticle also affects its stability and timescale of response to irradiation. Thus, for a reliable comparison between a theory and an experiment (as well as among different experiments) it is important to ensure that all such parameters are well controlled.

## 5. ACKNOWLEDGMENTS

The financial support from the Czech Ministry of Education, Youth, and Sports (grants No. LTT17015, LM2018114, and No. EF16_013/0001552) is gratefully acknowledged.

## REFERENCES

[1] Rossbach, J., Schneider, J. R. and Wurth, W., "10 years of pioneering X-ray science at the Free-Electron Laser FLASH at DESY," Phys. Rep. **808**, 1 (2019).
[2] Bostedt, C., Boutet, S., Fritz, D. M., Huang, Z., Lee, H. J., Lemke, H. T., Robert, A., Schlotter, W. F., Turner, J. J. and Williams, G. J., "Linac Coherent Light Source: The first five years," Rev. Mod. Phys. **88**(1), 015007 (2016).
[3] Owada, S., Togawa, K., Inagaki, T., Hara, T., Tanaka, T., Joti, Y., Koyama, T., Nakajima, K., Ohashi, H., Senba, Y., Togashi, T., Tono, K., Yamaga, M., Yumoto, H., Yabashi, M., Tanaka, H. and Ishikawa, T., "A soft X-ray free-electron laser beamline at SACLA: The light source, photon beamline and experimental station," J. Synchrotron Radiat. **25**(1), 282–288 (2018).
[4] Rethfeld, B., Ivanov, D. S., Garcia, M. E. and Anisimov, S. I., "Modelling ultrafast laser ablation," J. Phys. D.

Appl. Phys. **50**(19), 193001 (2017).
[5] Shugaev, M. V, Wu, C., Armbruster, O., Naghilou, A., Brouwer, N., Ivanov, D. S., Derrien, T. J. Y., Bulgakova, N. M., Kautek, W., Rethfeld, B. and Zhigilei, L. V., "Fundamentals of ultrafast laser-material interaction," MRS Bull. **41**(12), 960–968 (2016).
[6] Caricato, A. P., Luches, A. and Martino, M., "Laser fabrication of nanoparticles," [Handbook of Nanoparticles], Springer International Publishing, 407–428 (2015).
[7] Braun, M., Gilch, P. and Zinth, W., eds., [Ultrashort laser pulses in biology and medicine], Springer-Verlag, Berlin Heidelberg (2008).
[8] Sugioka, K. and Cheng, Y., "Ultrafast lasers—reliable tools for advanced materials processing," Light Sci. Appl. **3**(4), e149–e149 (2014).
[9] Bonse, J., Baudach, S., Krüger, J., Kautek, W. and Lenzner, M., "Femtosecond laser ablation of silicon-modification thresholds and morphology," Appl. Phys. A **74**, 19–25 (2002).
[10] Jiang, C.-W., Zhou, X., Lin, Z., Xie, R.-H., Li, F.-L. and Allen, R. E., "Electronic and Structural Response of Nanomaterials to Ultrafast and Ultraintense Laser Pulses," J. Nanosci. Nanotechnol. **14**(2), 1549–1562 (2014).
[11] Hau-Riege, S. P., [High-intensity X-rays - interaction with matter: processes in plasmas, clusters, molecules and solids], Willey-VCH Verlag, Weinheim, Germany (2011).
[12] Medvedev, N., "Femtosecond X-ray induced electron kinetics in dielectrics: application for FEL-pulse-duration monitor," Appl. Phys. B **118**(3), 417–429 (2015).
[13] Medvedev, N., Volkov, A. E. and Ziaja, B., "Electronic and atomic kinetics in solids irradiated with free-electron lasers or swift-heavy ions," Nucl. Instruments Methods Phys. Res. Sect. B Beam Interact. with Mater. Atoms **365**, 437–446 (2015).
[14] Keski-Rahkonen, O. and Krause, M. O., "Total and partial atomic-level widths," At. Data Nucl. Data Tables **14**(2), 139–146 (1974).
[15] Siders, C. W., Cavalleri, A., Sokolowski-Tinten, K., Tóth, C., Guo, T., Kammler, M., Hoegen, M. H. von, Wilson, K. R., Linde, D. von der and Barty, C. P. J., "Detection of nonthermal melting by ultrafast X-ray diffraction," Science **286**(5443), 1340–1342 (1999).
[16] Rousse, A., Rischel, C., Fourmaux, S., Uschmann, I., Sebban, S., Grillon, G., Balcou, P., Förster, E., Geindre, J. P., Audebert, P., Gauthier, J. C. and Hulin, D., "Non-thermal melting in semiconductors measured at femtosecond resolution.," Nature **410**(6824), 65–68 (2001).
[17] Medvedev, N., Tkachenko, V., Lipp, V., Li, Z. and Ziaja, B., "Various damage mechanisms in carbon and silicon materials under femtosecond x-ray irradiation," 4open **1**, 3 (2018).
[18] Medvedev, N., Babaev, P., Chalupský, J., Juha, L. and Volkov, A. E., "An interplay of various damage channels in polyethylene exposed to ultra-short XUV/x-ray pulse," Phys. Chem. Chem. Phys. **23**(30), 16193–16205 (2021).
[19] Medvedev, N., Voronkov, R. and Volkov, A. E., "Metallic water: Transient state under ultrafast electronic excitation," J. Chem. Phys. **158**(7), 074501 (2023).
[20] Medvedev, N., "Modeling warm dense matter formation within tight binding approximation," Proc. SPIE - Int. Soc. Opt. Eng. **11035** (2019).
[21] Medvedev, N., "Electronic nonequilibrium effect in ultrafast-laser-irradiated solids" (2023).
[22] Medvedev, N. and Milov, I., "Electron-phonon coupling in metals at high electronic temperatures," Phys. Rev. B **102**(6), 064302 (2020).
[23] Tully, J. C., "Molecular dynamics with electronic transitions," J. Chem. Phys. **93**(2), 1061 (1990).
[24] Mehl, M. J. and Papaconstantopoulos, D. A., "NRL transferable Tight-Binding parameters periodic table: http://esd.cos.gmu.edu/tb/tbp.html."
[25] Koskinen, P. and Mäkinen, V., "Density-functional tight-binding for beginners," Comput. Mater. Sci. **47**(1), 237–253 (2009).
[26] Akhmetov, F., Medvedev, N., Makhotkin, I., Ackermann, M. and Milov, I., "Effect of Atomic-Temperature Dependence of the Electron-Phonon Coupling in Two-Temperature Model," Materials (Basel). **15**(15), 5193 (2022).
[27] Medvedev, N. and Milov, I., "Nonthermal phase transitions in metals," Sci. Rep. **10**, 12775 (2020).
[28] Medvedev, N. and Milov, I., "Electron-Phonon Coupling and Nonthermal Effects in Gold Nano-Objects at High Electronic Temperatures," Materials (Basel). **15**(14), 4883 (2022).
[29] Medvedev, N., Kuglerová, Z., Makita, M., Chalupský, J. and Juha, L., "Damage threshold in pre-heated optical materials exposed to intense X-rays," Opt. Mater. Express **13**(3), 808 (2023).